\documentclass[]{article}

\usepackage{arxiv}

\usepackage[utf8]{inputenc} 
\usepackage[T1]{fontenc}    
\usepackage{hyperref}       
\usepackage{url}            
\usepackage{booktabs}       
\usepackage{amsfonts}       
\usepackage{nicefrac}       
\usepackage{microtype}      
\usepackage{graphicx}
\usepackage{natbib}
\usepackage{doi}
\usepackage{arydshln}
\usepackage{amsmath}
\usepackage{subcaption}
\usepackage{multirow}
\usepackage{authblk}
\usepackage{bm}

\usepackage[table]{xcolor}

\definecolor{myDarkGreen}{rgb}{0.1, 0.6, 0.1}
\definecolor{myLightGreen}{rgb}{0.7, 1.0, 0.7}
\definecolor{myYellow}{rgb}{1.0, 1.0, 0.8}
\definecolor{myRed}{rgb}{1.0, 0.8, 0.8}

\def\bSig\mathbf{\Sigma}

\usepackage{xurl}
\def\mydefb#1{\expandafter\def\csname b#1\endcsname{{\bm{#1}}}}
\def\mydefallb#1{\ifx#1\mydefallb\else\mydefb#1\expandafter\mydefallb\fi}
\mydefallb ABCDEFGHIJKLMNOPQRSTUVWXYZabcdeghijklnopqrstuvwxyz\mydefallb

\def\mydefgreek#1{\expandafter\def\csname b#1\endcsname{\text{\boldmath$\mathbf{\csname #1\endcsname}$}}}
\def\mydefallgreek#1{\ifx\mydefallgreek#1\else\mydefgreek{#1}%
	\lowercase{\mydefgreek{#1}}\expandafter\mydefallgreek\fi}
\mydefallgreek {beta}{Gamma}{Delta}{psi}{theta}{Psi}{zeta}{epsilon}{Omega}{etaex}{Theta}{ell}{Iota}{Lambda}{kappa}{mu}{nu}{Xi}{Pi}{rho}{Sigma}{tau}\mydefallgreek

\newtheorem{assumption}{Assumption}

\title{Informed Burn-In Decisions in RAR: \\ Harmonizing Adaptivity and Inferential Precision Based on Study Setting
}

\author[1]{Lukas Pin\textsuperscript{*}}
\author[1]{Stef Baas}
\author[1]{Gianmarco Caruso}
\author[1]{David S. Robertson}
\author[1]{Sofía S. Villar}

\affil[1]{Efficient Study Design Group, MRC Biostatistics Unit, University of Cambridge, Cambridge, UK}
\affil[*]{\textit{Corresponding author:} \href{mailto:lukas.pin@mrc-bsu.cam.ac.uk}{lukas.pin@mrc-bsu.cam.ac.uk}}

\renewcommand{\shorttitle}{\textit{Informed Burn-In Decisions in RAR}}

\hypersetup{
pdftitle={Informed Burn-In Decisions in RAR},
pdfsubject={biostatistics, RAR},
pdfauthor={Lukas Pin, Stef Baas, David Robertson, Sofia Villar},
pdfkeywords={Clinical trials; Design's reactiveness; Patient benefit; Power; Type-I error rate control},
}

\begin{document}
\maketitle

\begin{abstract}
Response-Adaptive Randomization (RAR) is recognized for its potential to deliver improvements in patient benefit. However, the utility of RAR is contingent on regularization methods to mitigate early instability and preserve statistical integrity. A standard regularization approach is the ``burn-in” period, an initial phase of equal randomization before treatment allocation adapts based on accrued data.
The length of this burn-in is a critical design parameter, yet its selection remains unsystematic and improvised, as no established guideline exists. A poorly chosen length poses significant risks: one that is too short leads to high estimation bias and type-I error rate inflation, while one that is too long impedes the intended patient and power benefits of using adaptation. The challenge of selecting the burn-in generalizes to a fundamental question: what is the statistically appropriate timing for the first adaptation?
We introduce the first systematic framework for determining burn-in length. This framework synthesizes core factors—total sample size, problem difficulty, and two novel metrics (reactivity and expected final allocation error)—into a single, principled formula. Simulation studies, grounded in real-world designs, demonstrate that lengths derived from our formula successfully stabilize the trial. The formula identifies a ``sweet spot” that mitigates type-I error rate inflation and mean-squared error, preserving the advantages of higher power and patient benefit. This framework moves researchers from conjecture toward a systematic, reliable approach.
\end{abstract}


\section{Introduction} \label{s:intro}

Adaptive designs introduce flexibility to clinical trials by allowing modifications based on accumulating data, thereby yielding more efficient and patient-oriented studies \citep{PallmannEtAl2018}. Response-Adaptive Randomization (RAR) is one such design where allocation probabilities change according to observed outcomes, typically to favor better-performing arms and/or increase statistical power.



Despite the potential advantages of RAR, its implementation requires careful management to ensure statistical validity and operational stability. To this end, regularization methods—such as tuning, clipping, and the burn-in period—are almost universally employed. These techniques are essential for controlling statistical properties like the type-I error rate, mitigating bias, and preventing the premature convergence of allocation based on unreliable early data. While such methods are widely adopted, a critical gap exists: as highlighted by \cite{Du2017Regularization}, there is limited guidance on how to systematically select or calibrate them for a given trial design.
The lack of clear, established guidance on how to effectively regularize RAR designs represents a high practical barrier to their optimal use in cases where they are most clinically relevant.

This paper focuses on the most common yet least explored of these techniques: the burn-in period. This is an initial phase where allocation follows a fixed randomization scheme, typically equal allocation, before switching to a response-adaptive rule. Its widespread use is underscored by a recent systematic review, which found that 88\% of RAR trials included a burn-in phase \citep{Wilson2025RARReview}. However, the choice of its length is often not explicitly justified. A burn-in that is too short may fail to stabilize the trial against early data fluctuations, while one that is too long may undermine potential power and/or patient benefits of the adaptive design. This reflects a fundamental trade-off between exploration and exploitation that is central to the design's success. While recent work by \cite{Tang2025} has begun to analyze the effect of the burn-in period on operating characteristics for the specific case using the Bayesian RAR (BRAR), a specific recommendation and generalized framework on how to choose the length of the burn-in period remains absent.

Here we address this gap by introducing the first systematic framework for determining the burn-in length in two-arm trials with binary outcomes. We analyze the fundamental components governing this decision: total sample size, intrinsic problem difficulty (standardized treatment effect), and two novel metrics—design reactiveness and expected final allocation error. These metrics quantify the inherent exploration-exploitation trade-off. The framework culminates in a single, principled formula (Equation \ref{eq:burnin_formula}) that synthesizes these distinct factors, offering a practical tool for researchers.


The paper is structured as follows: Section \ref{sec:notation} establishes the notation and RAR designs evaluated. Section \ref{s:metric} systematically develops our framework, introducing and analyzing the impact of the standardized treatment effect, sample size, reactiveness, and allocation error. Section \ref{s:method} presents our culminating formula and discusses its practical, adaptive implementation. Section \ref{s:simu} validates our approach in simulation studies based on two real-world trials. Finally, Section \ref{s:discuss} discusses the implications, limitations, and future research directions.

\section{Notation and considered designs}\label{sec:notation}

We first define the trial setting. Consider a trial with a total of $n$ patients and two arms: a control ($k=0$) and a treatment ($k=1$). Let $Y_{ki}$ be the potential outcome for patient $i$ on arm $k$, and $a_{ki}$ be the allocation indicator, where $\sum_{k=0}^{1}a_{ki}=1$ for all $i=1, \dots, n$. We assume patients are enrolled sequentially and outcomes are observed without delay. Although this might not align with all practical scenarios—where patients might enter trials in cohorts or where outcomes are delayed—we maintain these assumptions to preserve analytical tractability and clarity. Such assumptions are consistent with traditional RAR literature, which often avoids these complexities to enhance the understanding of methodological limits and potential gains over standard practices.

For the primary analysis, we consider binary endpoints where $Y_{ki} \sim Bern(p_k)$, with $p_k$ being the unknown response probability for arm $k$. For a trial with $K>1$ treatment arms, things might be even more complex since more treatment effects have to be taken into account. The multi-arm case is discussed in Section \ref{s:discuss}.

In the following Sections we compare some commonly known and novel RAR designs to an patient benefit benchmark design and equal randomization targeting equal allocation:

\begin{itemize}
    \item \textbf{Equal Randomisation (ER):} This term refers to designs that enforce \textit{equal allocation} ($n_0=n_1$). It represents the non-adaptive baseline in our framework. As our simulations do not involve time trends, our ER results can be interpreted as representative of any design that ensures a final 1:1 balance, such as the \textit{permuted block design}, \textit{random allocation rule}, or \textit{big stick design}. For more information and other designs see \cite{Berger2021}.
    
    \item \textbf{Patient Benefit Benchmark Design (PBB):} Serving as a theoretical benchmark for patient benefit, this design knows the ground truth and allocates all subsequent patients to the superior arm following the burn-in period. We use the PBB—even though it needs no learning phase—to precisely analyze the burn-in's isolated impact on key operating characteristics, particularly patient benefit.
    
    \item \textbf{Bayesian Response-Adaptive Randomization (BRAR (U))}: The untuned version of BRAR stems from \cite{Thompson1933}. Also known as \textbf{Thompson Sampling}, this unregularized Bayesian design allocates the next patient to arm 1 with a probability equal to the current posterior probability that arm 1 is superior, i.e., $P(p_1 > p_0 \mid \text{data})$. We use a non-informative Beta(1,1) prior.
    
    \item \textbf{Bayesian Response-Adaptive Randomization (BRAR (T))}: The tuned version of BRAR was proposed by \cite{Thall2015}. This is a regularized Bayesian design. It tunes the allocation probabilities in a way that for earlier patients the probabilities closer towards 0.5 than the ones of BRAR (U). 
    
    \item \textbf{Neyman Allocation for the Wald test ($N_1$)}: Neyman allocation \citep{Neyman1934} is derived to maximize statistical power for the Wald test ($Z_1$). The allocation proportion is targeted using the Efficient Randomized-Adaptive Design (ERADE) \citep{Hu2009}.
    
    \item \textbf{Neyman Allocation for the score test ($N_0$)}: This is a novel power-optimizing allocation derived in \cite{pin2025revisitingoptimalallocationsbinary}, targeted using ERADE. This proportion is specifically designed to maximize statistical power when the final analysis is conducted using the more robust score test ($Z_0$) preventing type-I error inflation.
    
    \item \textbf{RSHIR Allocation for the Wald test ($R_1$)}: This is the classical RSHIR allocation \citep{Rosenberger2001}, targeted with ERADE. This allocation proportion is derived to optimize for patient benefit by minimizing the expected number of failures in the trial, subject to a constraint on the variance of the Wald test ($Z_1$).
    
    \item \textbf{RSHIR Allocation for the score test ($R_0$)}: This is a novel patient-benefit-optimizing allocation from \cite{pin2025revisitingoptimalallocationsbinary}, targeted with ERADE. It minimizes the expected number of failures subject to a constraint on the variance of the score test ($Z_0$).
    
    \item \textbf{The Play the Winner Rule (PTW)} \citep{Zelen1969}: This is a deterministic allocation rule. After a success on an arm, the next patient is allocated to the same arm. After a failure, the next patient is allocated to the other arm.
    
    \item \textbf{The Randomized Play the Winner Rule (RPTW)} \citep{Wei1978}: A ``success-driven" urn-based design. An urn contains balls of two types (e.g., red and blue). Initially it contains one ball of each type. The next patient's allocation is determined by drawing a ball from the urn. When a success is observed on an arm (e.g., red), a red ball is added to the urn. If a failure is observed, a blue ball is added. 
\end{itemize}

The list of designs evaluated is representative, not exhaustive. The procedure we will define in the following sections is general and valid for any RA(R) algorithm, including those yet to be developed. We use ER and the PBB design as fundamental benchmarks to frame the analysis, as they represent the non-adaptive baseline and the theoretical ideal for patient benefit, respectively. The two Neyman allocations $N_1$ and $N_0$ are theoretically (and asymptotically for $N_1$) optimal in terms of power for the Wald and score test, respectively, but they do not always achieve this in finite samples \citep{pin202511powerfulcarefuldetermination}.

In the next section we investigate what influences the burn-in length $b$, where $b$ is the number of patients allocated to each arm before we start adapting.

\section{What influences the burn-in length and how?}\label{s:metric}

\subsection{Standardized treatment effect}\label{s.standardisedTE}

\paragraph{Treatment effect}
We define the treatment effect as the simple difference between the experimental and control response probabilities: $p_1-p_0$. A smaller treatment effect (i.e., a small difference) makes it more difficult to correctly identify the superior arm, even when one truly exists. This scenario creates greater initial uncertainty, suggesting that a longer burn-in period may be beneficial to ensure adequate exploration of all arms before adaptation begins. While our analysis focuses on this simple difference, other measures like the log relative risk or log odds ratio could be used, but they would can alter the RAR design and the operating characteristics of the final inference method \citep{Pin_Deming2024}, see Section \ref{s:discuss} for more details.

\paragraph{Arm-specific response variances}
The more variable the response of an arm, the more observations it requires to reliably estimate its expected outcome. Ideally, when arm-specific response variances are high, we might want to guarantee that the arms are ``sufficiently" explored through a longer burn-in period. 

Thus, rather than considering the simple treatment effect, we suggest considering a standardized treatment effect, i.e.,
\begin{equation}
\label{eq:standardTreatDiff}
\delta = \frac{|p_1 - p_0|}{\sqrt{p_0(1-p_0) + p_1(1-p_1)}},
\end{equation}
which measures the difference in response probabilities scaled by the total variability in the two arms. The metric is unbounded and approaches infinity as the treatment effect $|p_1-p_0|$ approaches $1$ while both arm-specific variances approach $0$ (e.g., as $p_0 \to 0$ and $p_1 \to 1$, or vice-versa).
The quantity is only undefined in degenerate cases where both responses are deterministic (i.e.\ $p_0, p_1 \in \{0,1\}$). In these situations, we define
\[
\delta =
\begin{cases}
0, & \text{if } p_1 = p_0,\\[4pt]
\infty, & \text{if } p_1 \neq p_0.
\end{cases}
\]
The standardized effect $\delta$ therefore lies in the range $[0, \infty]$. The lower bound $\delta = 0$ occurs when there is no treatment effect, i.e., $p_1 = p_0$, provided that $p_0$ and $p_1$ are not both 0 or both 1. The metric is technically undefined if $p_0=p_1=0$ or $p_0=p_1=1$, as this results in an indeterminate $0/0$ form. A larger $\delta$ signifies a clearer, more easily detectable difference between the arms relative to their variability.

The standardized treatment effect accounts for the fact that response probabilities closer to $0.5$ are associated to a higher variability in the arm response than response probabilities which are close to the boundaries $0$ and $1$. For example, let us consider two scenarios, each one characterized by different pairs of response probabilities, i.e. $p_0=0.7,\,p_1=0.9$ (first scenario) and $p_0=0.4,\,p_1=0.6$ (second scenario). Although the treatment effect is $0.2$ in both scenarios, the second scenario is associated with higher variability than the first one and, thus, we would require a higher exploration (i.e. a higher burn-in) under this scenario. Indeed, the higher total variability of the responses of the second scenario is reflected in a smaller value of the standardized treatment effect ($\delta\approx0.29$) than the first scenario ($\delta\approx0.37$). 

\begin{figure}[]
    \centering
    \includegraphics[width=0.95\linewidth]{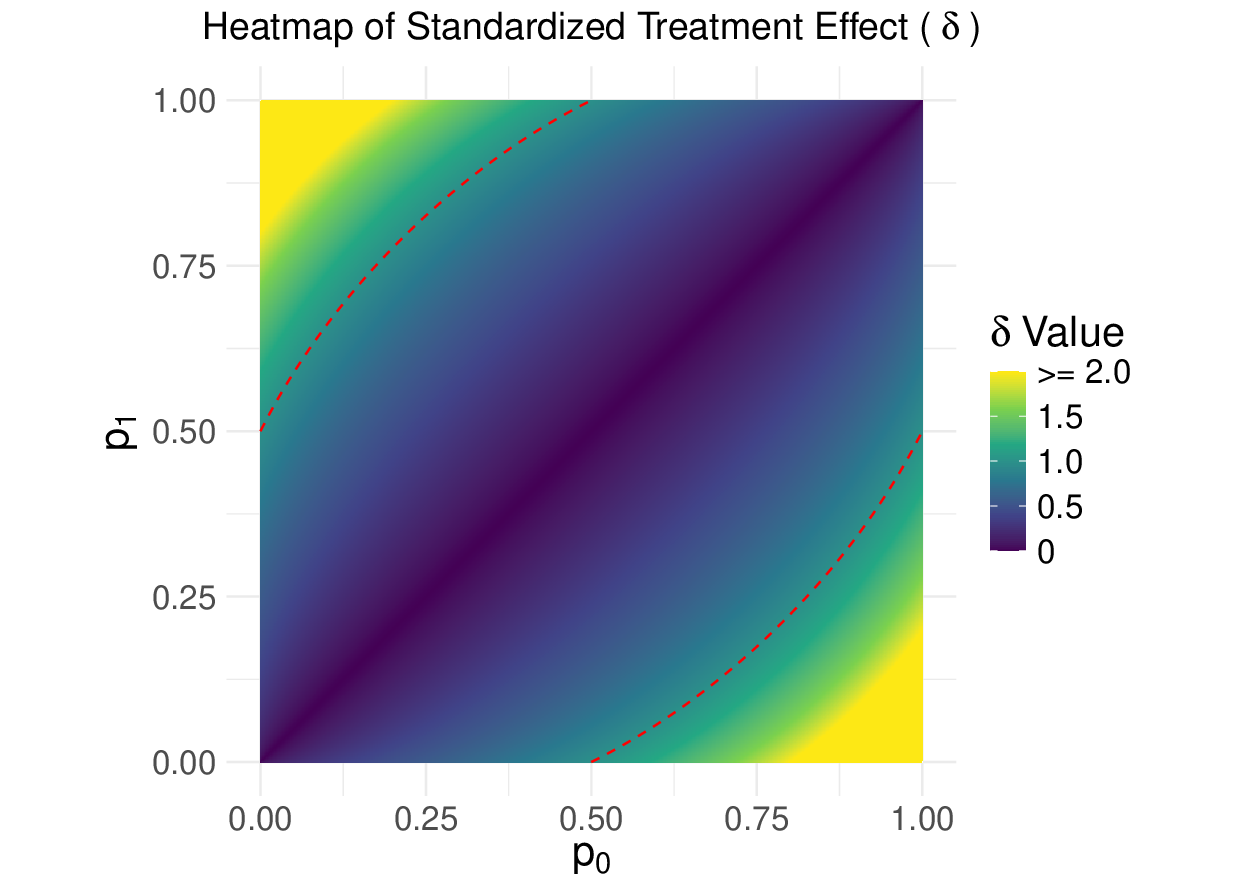}
    \caption{Heatmap of the standardized treatment effect ($\delta$) across the full parameter space. The dashed red contour marks $\delta=1$. The color scale is capped at 2 for visualization, as $\delta$ approaches infinity in the top-left ($p_0 \to 0, p_1 \to 1$) and bottom-right ($p_0 \to 1, p_1 \to 0$) corners.}
    \label{fig:delta}
\end{figure}

While $\delta$ is theoretically unbounded (ranging from $0$ to $\infty$), Figure \ref{fig:delta} reveals that $\delta < 1$ for the majority of the parameter space. The red contour ($\delta=1$) highlights the boundary of this region. A key observation is that $\delta$ can only exceed 1 if the absolute treatment difference, $|p_1 - p_0|$, is greater than 0.5. This minimum difference is only sufficient at the boundaries (e.g., $p_0=0, p_1=0.5$). As the probabilities move away from 0 or 1 and towards 0.5, the treatment difference required to maintain $\delta=1$ increases, as shown by the red curve's shape. This effect is most pronounced when one probability equals 0.5. In that case ($p_0 = 0.5$ or $p_1 = 0.5$), $\delta$ reaches its maximum possible value of 1, regardless of how extreme the other probability becomes.

In practice, $\delta$ can be calculated using the response probabilities specified for the trial's power analysis. Sample size determination typically requires an assumed treatment effect and a baseline success probability, $p_0$, often derived from previous studies. If $p_0$ is specified as a range rather than a single value, one could use the midpoint of this range. Alternatively, one might select the $p_0$ value within the range that results in the highest standardized treatment effect, although this would represent a more optimistic scenario for detection.

\paragraph{Impact on burn-in}

A smaller expected standardized treatment effect  suggests that a larger burn-in period is required. This is because a small $\delta$ indicates that the difference between the arms' response probabilities is difficult to detect. This difficulty can be caused by either a small true treatment effect (a small numerator) or high response variability (a large denominator). The $\delta$ metric effectively captures both of these challenges in a single variable.

\subsection{Impact of sample size on burn-in}\label{sec:samplesize}

\begin{figure}[]
    \centering
    \includegraphics[width=0.95\linewidth]{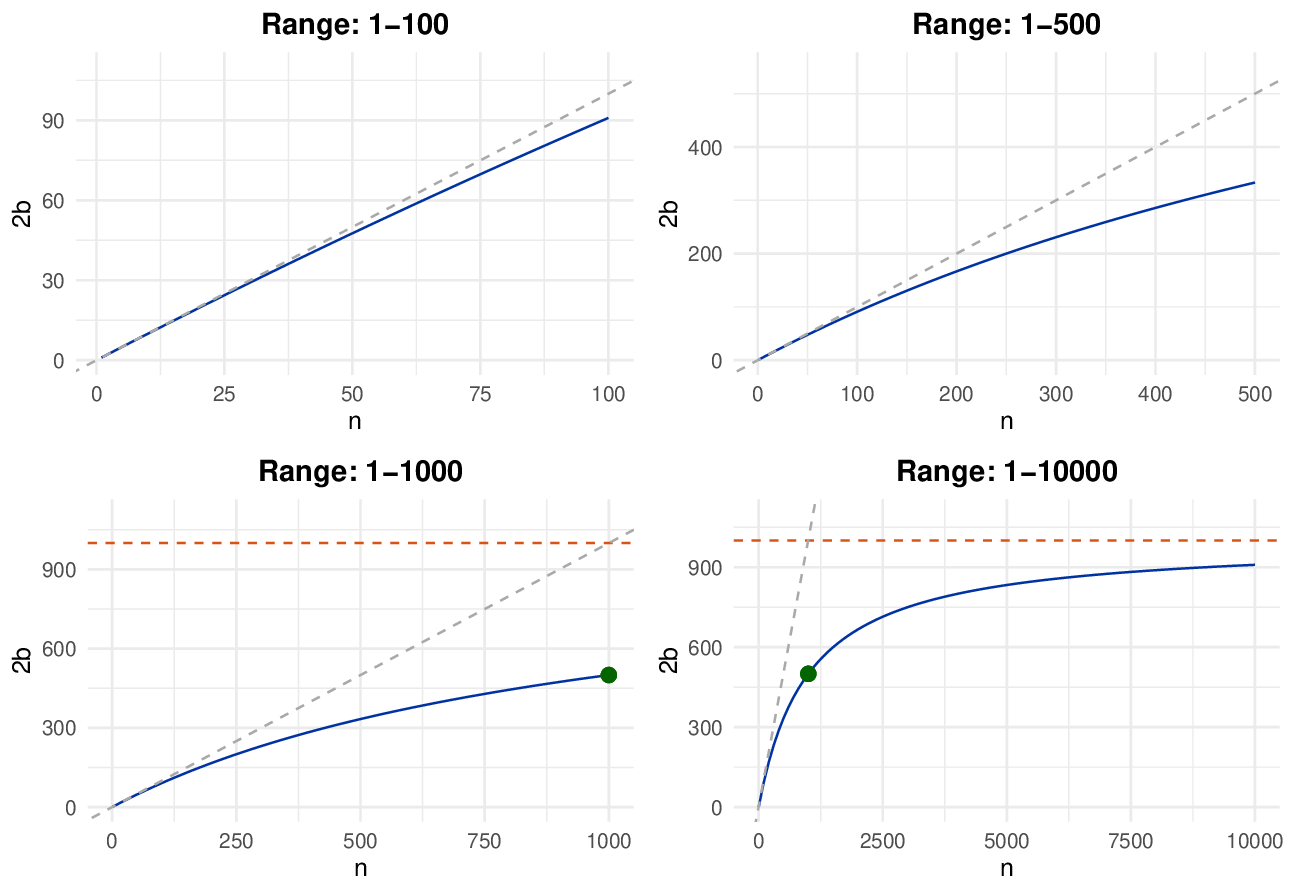}
    \caption{Total available burn-in proportion (largest $2b$ possible) based on sample size ($n$) and $\frac{n \cdot n_{1/2}}{n+n_{1/2}}$ with $n_{1/2}=1000$. The gray dashed line reflects the largest burn-in budget upper bound ($2b=n$).}
    \label{fig:2b_curve}
\end{figure}

The burn-in serves two primary functions: (1) to stabilize estimators at the beginning of the trial, and (2) to improve inferential properties, such as type-I error rate and power. We assume that for a given treatment effect, these objectives are typically met in a standard, non-adaptive trial (using only ER) once the total sample size $n$ is sufficiently large. We posit that this same level of stability and performance can be achieved within the burn-in phase of our adaptive trial, provided the total burn-in size ($2b$) is large enough. 
Once this sufficient burn-in size is reached, additional burn-in patients provide diminishing or no benefit. Consequently, the proportion of patients allocated to the burn-in period ($BP = 2b/n$) should decrease as the total trial size n increases
To ensure BP decreases appropriately, we define a non-linear term
\begin{equation}\label{eq:samplesizef}
    \frac{n \cdot n_{1/2}}{n+n_{1/2}},
\end{equation}
where $n_{1/2}$ is a sample size parameter that defines the non-linear decrease.
This term functions as an available sample size for the burn-in, which grows more slowly than $n$. By setting the sample size parameter $n_{1/2}$ equal to $1000$, we achieve a largest total burn-in ($2b$) of 500 for a trial with $n=1000$, see Figure \ref{fig:2b_curve}. 

\subsection{Reactiveness of a RAR design}\label{sec.reactiveness}

Here, we present an operational definition for reactiveness of a RAR design. With reactiveness we describe a characteristic of adaptive designs that has been referred to as `aggressiveness' without formally defining it, see e.g. \cite{VillarBowdenWason2018, Viele2025, BAAS2025108207, BaasPinVillarRosenberger2025}.

The allocation proportion to the experimental treatment after allocation of participant~$i$ in the trial can be calculated based on the current patient allocation, $n(i)=(n_{0}(i),n_{1}(i))$, as $n_{1}(i)/i$.
The following assumption is made:
 \begin{assumption}\label{assump:convg}
     The RA(R) procedure is such that~$\lim_{i\rightarrow\infty}n_{1}(i)/i=\rho$ exists for almost every~$\{ p_0,p_1 \}\in[0,1]^2$.
 \end{assumption}
This assumption holds for most RA(R) procedures.
Note that ``almost every~${\{ p_0,p_1 \}}\in[0,1]^2$'' excludes the case~$p_0=p_1$ as this line has no (Lebesgue) measure in~$[0,1]^2$.
This detail enforces that Assumption~\ref{assump:convg} holds for procedures such as unregularized  Bayesian RAR~(i.e, Thompson sampling) where $n_1(i)/i$ may converge in distribution to an uniform random variable when~$p_0=p_1$~\citep[See, e.g., Proposition~1 in][showing this result for the normal-normal model]{NEURIPS2020_6fd86e0a}.
For RA(R) procedures such as the randomized play-the-winner rule (RPTW),  BRAR or targeting optimal proportions,~$\rho$ is known. 
If the theoretical limit is unknown we can estimate the limiting proportions through simulations. 

Given the limiting proportion~$\rho$ for the RA(R) procedure, we can estimate the expected speed of convergence of~$n_{1}(i)/i$ to~$\rho$.
We assume for~$c>0$ that
\begin{equation}
    n_{1}(i)/i\approx \rho+ (1/2-\rho)i^{-c}.\label{approxrho}
\end{equation}
The approximation~\eqref{approxrho} is based on the assumption that the smallest burn-in of two patients per arm is employed, $\lim_{i\rightarrow\infty}n_{1}(i)/i=\rho,$ and furthermore $n_{1}(i)/i=\rho+O(n^{-c})$ for~$c>0$~(see, e.g.,~\citet{yi2018response}, for a similar assumption). From~\eqref{approxrho} we obtain~\begin{equation}
    c\approx \begin{cases}
        -\log(|n_{1}(i)/i-\rho|/|1/2-\rho|)/\log(i),\indent&\text{if~$\rho\neq 1/2$,}\\
        0,&\text{otherwise.}
    \end{cases}\label{eqn:speed}
\end{equation}
While~$c$ can be determined exactly from one realization (e.g.,~$n_{1}(n)/n$) of an allocation proportion when Equation ~\eqref{approxrho} is a perfect fit~(i.e., for deterministic procedures with perfect knowledge), it is likely the case for practical RAR procedures that~\eqref{approxrho} is only an approximation, hence we will estimate the geometric slope~$c$ from the complete trial realization with total trial size~$n$, i.e., using~$$\hat{c}(\rho)=\frac{1}{n}\sum_{i=2}^{n} -\log(|n_{1}(i)/i-\rho|/|1/2-\rho|)/\log(i)\cdot\mathbb{I}(\rho\neq 1/2).$$ 
In the above, the exclusion of~$i$=1 follows from the fact that~$\log(1)=0.$
The integral of the geometric slope, independent of the parameters $p_0$ and $p_1$, equals~$\int_{[0,1]^2}c\cdot d\bp$ (where $\bp$ is the vector $(p_0,p_1)$) and can be estimated using Monte Carlo sampling:
\begin{equation}\label{eq:slopeMC}
    \tilde{r}(\rho)=\frac{1}{n_{\text{sim}}}\sum_{m=1}^{n_{\text{sim}}}\hat{c}_m(\rho_m)
\end{equation}
where~$\rho_m$ is the target proportion following from parameter vector~$\bp_{m}\sim U([0,1]^2)$ for all~$m\in\{1,\dots, n_{\text{sim}}\}.$

In practice, it was found that the proportion~$n_{1}(i)/i$ can get stuck around zero or one for very reactive RAR procedures, even if this is not the limit proportion. This is because, in such settings where most participants are allocated to one treatment, little is learned about the actual differences in the treatment groups. Due to this, it may be that at the end of the trial the difference~$|n_1(n)/n-\rho|$ is larger than~$|n_1(n)/n-1/2|$ and hence we get~$\hat{c}_m<0$. To this end, we estimate the reactiveness parameter, denoted~$r$, as
\begin{equation}
    \label{eq:reactivness}
    r = \max(0, \tilde{r}(\rho), \tilde{r}(1), \tilde{r}(0)),
\end{equation}
i.e., the reactiveness is restricted to be nonnegative, and defined as the maximum rate at which the allocation proportion goes to zero, one, or the theoretical limit.
For RAR procedures, the reactiveness parameter $r$ is expected to take on values between $0$ and $0.5$, this is because sampling with a fixed biased coin design~(i.e., where the target proportion is known) leads to an asymptotic convergence rate of~$1/2$ (indicated by the central limit theorem). For RAR procedures, i.e. procedures that are learning the target proportion sequentially during the trial, this convergence will likely be slower.  For instance, \citet{Hu2006} show that the central limit theorem holds with slower rates than 1/2 for RPTW and rates equal to 1/2 for RAR based on sequential estimation such as ERADE. We note that that these are asymptotic rates, and when the reactiveness is estimated based on a finite horizon as in~\eqref{eqn:speed}, the estimated reactiveness parameter will likely be lower. This is because the allocated proportion under a RAR procedure may not have converged to the asymptotic limit proportion.

Table \ref{tab:generallagressivness} reports comparisons of the reactiveness parameters across several RAR procedures~(listed in Section~\ref{sec:notation}) for $n = 200, 500, 1000,$ and $2000$. The slope defined in Equation~\eqref{eq:slopeMC} is computed independently of the parameters $p_0$, $p_1$. We observe that the resulting estimates change over the different choices of~$n$.

\begin{table}[htbp]
    \centering
    \caption{Estimated reactiveness parameters ($r$ and $\epsilon_\rho$) and resulting burn recommendation ($b$) and burn-in proportion ($BP$), independent of parameters~$p_0,p_1$, for trial sizes~$n\in\{200,500,1000,2000\}$ based on $n_{\text{sim}}=1000$ simulations.}
    \label{tab:generallagressivness}
    \small\begin{tabular}{llccccc}
\hline
$n$ & Design & $r$ (x100) & $\epsilon_{\rho}$ (x100) & $r + \epsilon_{\rho}$ (x100) & $b$ & \multicolumn{1}{c}{$BP$ (x100)} \\ 
\hline
\nopagebreak 200 & \nopagebreak ER  & 0.00 +/- 0.00 & 0.00 +/- 0.00 & 0.00 +/- 0.00 & - & - \\
 & \nopagebreak PBB  & 65.77 +/- 0.00 & 0.00 +/- 0.00 & 65.77 +/- 0.00 & 65 +/- 0.82 & 64.94 +/- 0.82 \\
 & \nopagebreak BRAR (U)  & 34.25 +/- 1.17 & 0.83 +/- 0.31 & 35.09 +/- 1.19 & 51 +/- 0.96 & 50.13 +/- 0.96 \\
 & \nopagebreak BRAR (T)  & 9.72 +/- 0.41 & 0.36 +/- 0.12 & 10.08 +/- 0.41 & 30 +/- 1.29 & 29.55 +/- 1.29 \\
 & \nopagebreak $N_0$  & 18.91 +/- 1.09 & 1.53 +/- 0.18 & 20.43 +/- 1.10 & 37 +/- 1.39 & 36.63 +/- 1.39 \\
 & \nopagebreak $N_1$  & 13.62 +/- 1.03 & 32.41 +/- 1.06 & 46.03 +/- 1.21 & 54 +/- 1.26 & 53.64 +/- 1.26 \\
 & \nopagebreak $R_0$  & 19.48 +/- 1.12 & 1.18 +/- 0.13 & 20.65 +/- 1.12 & 37 +/- 1.43 & 36.37 +/- 1.43 \\
 & \nopagebreak $R_1$  & 25.12 +/- 1.15 & 16.97 +/- 1.12 & 42.09 +/- 1.48 & 54 +/- 1.11 & 53.91 +/- 1.11 \\
 & \nopagebreak PTW  & 36.15 +/- 1.52 & 1.29 +/- 0.15 & 37.44 +/- 1.49 & 52 +/- 1.08 & 51.88 +/- 1.08 \\
 & \nopagebreak RPW  & 25.41 +/- 1.29 & 1.79 +/- 0.21 & 27.20 +/- 1.25 & 43 +/- 1.25 & 42.30 +/- 1.25 \\
\rule{0pt}{1.7\normalbaselineskip} 500 & \nopagebreak ER  & 0.00 +/- 0.00 & 0.00 +/- 0.00 & 0.00 +/- 0.00 & -& - \\
 & \nopagebreak PBB  & 72.02 +/- 0.00 & 0.00 +/- 0.00 & 72.02 +/- 0.00 & 137 +/- 1.46 & 54.46 +/- 0.58 \\
 & \nopagebreak BRAR (U)  & 38.61 +/- 1.24 & 1.07 +/- 0.36 & 39.68 +/- 1.24 & 109 +/- 1.68 & 43.30 +/- 0.67 \\
 & \nopagebreak BRAR (T)  & 11.93 +/- 0.47 & 0.12 +/- 0.06 & 12.05 +/- 0.46 & 63 +/- 2.37 & 24.81 +/- 0.95 \\
 & \nopagebreak $N_0$  & 23.05 +/- 1.07 & 0.90 +/- 0.12 & 23.95 +/- 1.07 & 80 +/- 2.78 & 31.73 +/- 1.11 \\
 & \nopagebreak $N_1$  & 21.01 +/- 1.19 & 33.94 +/- 1.06 & 54.95 +/- 1.44 & 119 +/- 2.59 & 47.58 +/- 1.04 \\
 & \nopagebreak $R_0$  & 24.69 +/- 1.14 & 0.85 +/- 0.11 & 25.54 +/- 1.14 & 82 +/- 2.75 & 32.46 +/- 1.10 \\
 & \nopagebreak $R_1$  & 33.84 +/- 1.19 & 17.38 +/- 1.15 & 51.22 +/- 1.66 & 118 +/- 2.15 & 46.92 +/- 0.86 \\
 & \nopagebreak PTW  & 38.76 +/- 1.35 & 0.85 +/- 0.10 & 39.61 +/- 1.33 & 109 +/- 1.83 & 43.42 +/- 0.73 \\
 & \nopagebreak RPW  & 26.74 +/- 1.23 & 1.25 +/- 0.14 & 27.99 +/- 1.20 & 91 +/- 2.29 & 36.13 +/- 0.92 \\
\rule{0pt}{1.7\normalbaselineskip} 1000 & \nopagebreak ER  & 0.00 +/- 0.00 & 0.00 +/- 0.00 & 0.00 +/- 0.00 &- & - \\
 & \nopagebreak PBB  & 75.50 +/- 0.00 & 0.00 +/- 0.00 & 75.50 +/- 0.00 & 210 +/- 2.02 & 41.93 +/- 0.40 \\
 & \nopagebreak BRAR (U)  & 43.92 +/- 1.22 & 0.60 +/- 0.26 & 44.52 +/- 1.21 & 170 +/- 2.10 & 33.88 +/- 0.42 \\
 & \nopagebreak BRAR (T)  & 13.23 +/- 0.49 & 0.10 +/- 0.06 & 13.32 +/- 0.49 & 102 +/- 3.31 & 20.38 +/- 0.66 \\
 & \nopagebreak $N_0$  & 26.03 +/- 1.03 & 0.57 +/- 0.07 & 26.59 +/- 1.04 & 128 +/- 4.04 & 25.48 +/- 0.81 \\
 & \nopagebreak $N_1$  & 28.11 +/- 1.23 & 34.54 +/- 1.05 & 62.65 +/- 1.50 & 192 +/- 3.66 & 38.21 +/- 0.73 \\
 & \nopagebreak $R_0$  & 29.64 +/- 1.10 & 0.62 +/- 0.07 & 30.27 +/- 1.09 & 135 +/- 3.77 & 26.98 +/- 0.75 \\
 & \nopagebreak $R_1$  & 37.76 +/- 1.20 & 18.41 +/- 1.18 & 56.17 +/- 1.78 & 188 +/- 3.31 & 37.42 +/- 0.66 \\
 & \nopagebreak PTW  & 40.25 +/- 1.26 & 0.58 +/- 0.08 & 40.83 +/- 1.25 & 165 +/- 2.57 & 32.80 +/- 0.51 \\
 & \nopagebreak RPW  & 30.39 +/- 1.21 & 0.85 +/- 0.12 & 31.24 +/- 1.19 & 138 +/- 3.20 & 27.59 +/- 0.64 \\
\rule{0pt}{1.7\normalbaselineskip} 2000 & \nopagebreak ER  & 0.00 +/- 0.00 & 0.00 +/- 0.00 & 0.00 +/- 0.00 & - & - \\
 & \nopagebreak PBB  & 78.24 +/- 0.00 & 0.00 +/- 0.00 & 78.24 +/- 0.00 & 288 +/- 2.28 & 28.72 +/- 0.23 \\
 & \nopagebreak BRAR (U)  & 45.72 +/- 1.18 & 0.51 +/- 0.24 & 46.23 +/- 1.16 & 235 +/- 2.66 & 23.45 +/- 0.27 \\
 & \nopagebreak BRAR (T)  & 15.46 +/- 0.55 & 0.09 +/- 0.05 & 15.55 +/- 0.54 & 144 +/- 4.65 & 14.31 +/- 0.46 \\
 & \nopagebreak $N_0$  & 28.98 +/- 0.94 & 0.40 +/- 0.04 & 29.37 +/- 0.93 & 177 +/- 5.35 & 17.62 +/- 0.53 \\
 & \nopagebreak $N_1$  & 34.87 +/- 1.28 & 34.35 +/- 1.07 & 69.22 +/- 1.67 & 274 +/- 4.62 & 27.36 +/- 0.46 \\
 & \nopagebreak $R_0$  & 31.65 +/- 1.07 & 0.39 +/- 0.05 & 32.04 +/- 1.07 & 186 +/- 5.05 & 18.51 +/- 0.51 \\
 & \nopagebreak $R_1$  & 42.11 +/- 1.17 & 17.84 +/- 1.17 & 59.95 +/- 1.86 & 258 +/- 4.43 & 25.73 +/- 0.44 \\
 & \nopagebreak PTW  & 41.76 +/- 1.17 & 0.39 +/- 0.04 & 42.16 +/- 1.16 & 221 +/- 3.20 & 22.10 +/- 0.32 \\
 & \nopagebreak RPW  & 31.05 +/- 1.13 & 0.69 +/- 0.10 & 31.74 +/- 1.11 & 190 +/- 4.15 & 18.91 +/- 0.41 \\
\hline 
\end{tabular}

\end{table}

\subsubsection{Reactiveness in practice}

In practical settings, the treatment difference $(p_1 - p_0)$, the baseline response probability $p_0$, and the targeted sample size $n$ are typically given in advance, as they are determined by power considerations during trial planning. 
If only a range for $p_0$ is known we can choose the central value in this range or the value that yields the largest $\delta$. Thus, instead of evaluating reactiveness across randomly drawn parameter pairs $(p_0, p_1) \sim U([0,1]^2)$, we adapt the measure to reflect the specific planned trial characteristics.

For specific parameter configurations, we define reactiveness as
\begin{equation}
    \frac{1}{n_{\text{sim}}}\sum_{m=1}^{n_{\text{sim}}}
    \hat{c}_m(\rho),
\end{equation}
where $n$ denotes the planned trial size, $n_{\text{sim}}$ the number of simulated trials and $\rho$ the optimal proportion for the specific parameter combination. In this formulation, simulations are carried out using fixed parameter values $p_0$ and $p_1$, rather than sampling them from a uniform distribution.
This yields an reactiveness measure that reflects how the design behaves under the \emph{expected} conditions of the planned trial, rather than across a broad parameter space.

\begin{figure}
    \centering
    \includegraphics[width=\linewidth]{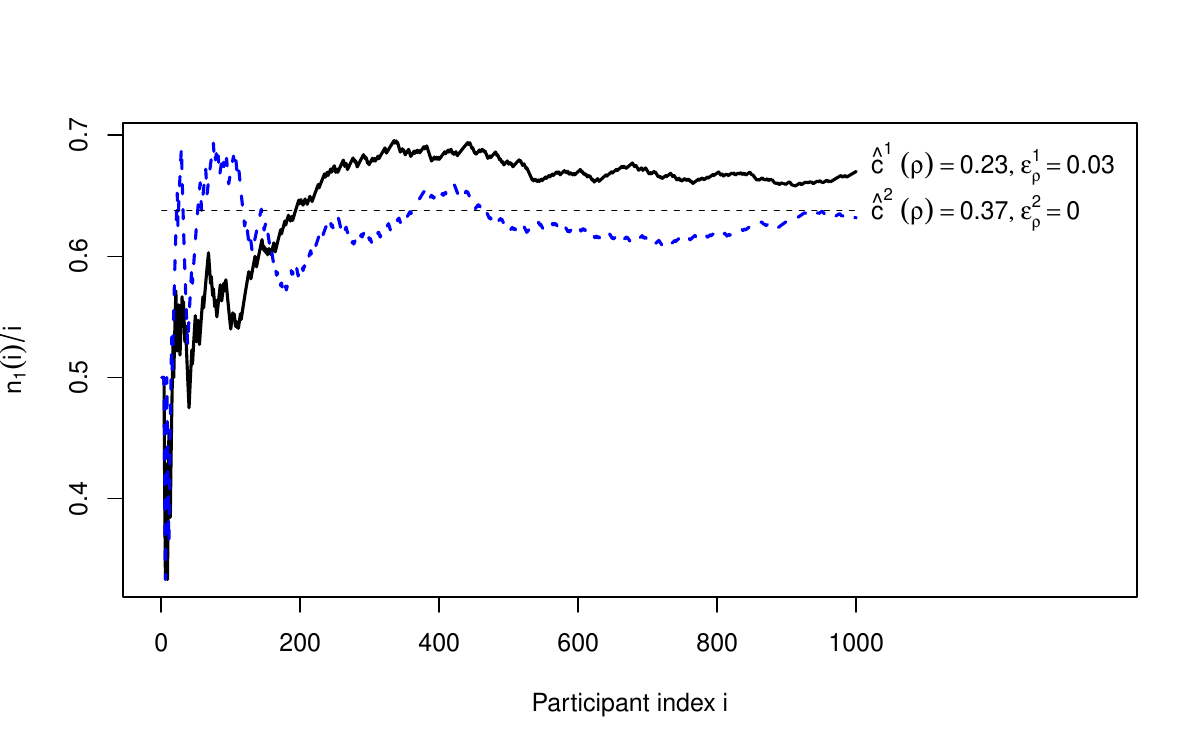}
    \caption{Two sampled paths~(1000 participants) for the $R_0$ design under~$p_0=0.6$, $p_1=0.8$ leading \hbox{to~$\rho=0.64$.} The solid path displays a slower trend than the dashed path and has not converged yet, ending up at a value higher than~$\rho$; hence, the estimated geometric slope is~$0.23$, while the error~$\epsilon^{1}_{\rho}$ is around 0.03. The dashed path has converged to a value around~$\rho$ and hence has a higher slope~$\hat{c}_2(\rho)=0.37$ than the solid path and error~$\epsilon^{2}_{\rho}=0.00$ as the final proportion is below~$\rho.$ }
    \label{fig:reactiveness}
\end{figure}

\paragraph{Impact on burn-in}
A design with high reactiveness requires a larger burn-in. This is because the algorithm moves away from a balanced allocation quickly, and a burn-in is necessary to prevent this rapid convergence from being based on misleading early estimates.
This need for a burn-in is further amplified if the design's limiting allocation ($\rho$) is far from 0.5. In such cases, the burn-in is not only protecting against premature convergence but also ensuring sufficient exploration before the design commits to a heavily imbalanced allocation. Therefore, the greatest need for a burn-in occurs when a design is both converging quickly and targeting an extreme allocation ($\rho$ far from 0.5).

\subsection{Expected final allocation error}\label{sec:finalerror}

The reactiveness measure $r$, defined in Section \ref{sec.reactiveness}, measures the speed of convergence and the extremeness of the limiting allocation, see Equation \eqref{eq:reactivness}. 
This new measure, $\epsilon_\rho$, addresses a different question: How accurately does the algorithm's final allocation match what it is supposed to do?

While $r$ evaluates the reactiveness of the allocation, it does not penalize an algorithm for converging to the wrong target (e.g., converging to 0.1 when the target $\rho$ is 0.8). We introduce $\epsilon_\rho$ to specifically evaluate the accuracy of the final allocation proportion, $(n_0,n_1)/n$, relative to its theoretical target $\rho$. The goal is to isolate this accuracy error from the separate concept of reactivity, thereby avoiding a double penalty for designs that are intentionally reactive, such as BRAR.

The error for a single trial, $\epsilon_{\rho,n}$, is defined by the following logic:
\begin{itemize}
    \item If $\rho \ge 0.5$, the target interval is $[0.5, \rho]$.
    \begin{itemize}
        \item If $n_1/n \in [0.5, \rho]$, $\epsilon_{\rho,n} = 0$.
        \item If $n_1/n > \rho$, $\epsilon_{\rho,n} = n_1/n - \rho$.
        \item If $n_1/n < 0.5$, $\epsilon_{\rho,n} = 0.5 - n_1/n$.
    \end{itemize}
    \item If $\rho < 0.5$, the target interval is $[\rho, 0.5]$.
    \begin{itemize}
        \item If $n_0/n \in [\rho, 0.5]$, $\epsilon_{\rho,n} = 0$.
        \item If $n_0/n < \rho$, $\epsilon_{\rho,n} = \rho - n_0/n$.
        \item If $n_0/n > 0.5$, $\epsilon_{\rho,n} = n_0/n - 0.5$.
    \end{itemize}
\end{itemize}
A key feature of this metric is its behavior under the null hypothesis (i.e., when $\rho = 0.5$). In this scenario, the target interval $[0.5, \rho]$ collapses to the single point $[0.5, 0.5]$. The definition correctly simplifies to penalize any deviation from a balanced 0.5 allocation, and it does so symmetrically. The error becomes $\epsilon_{0.5,n} = |n_1/n - 0.5|$.

We then report the \textit{Expected Final Allocation Error} $\epsilon_{\rho}$ by averaging $\epsilon_{\rho,n}$ across $n_{\text{sim}}$ simulations.
To illustrate, let us assume $\rho \ge 0.5$, for example $\rho = 0.6$. The target interval is $[0.5, 0.6]$.
\begin{itemize}
    \item If the final allocation lies in the interval, e.g. $n_1(n)/n = 0.55$, then $\epsilon_{\rho,n} = 0$.
    \item If the final allocation $n_1(n)/n = 0.7$ (an ``overshoot"), the error is $\epsilon_{\rho,n} = 0.7 - 0.6 = 0.1$.
    \item If the final allocation $n_1(n)/n = 0.3$ (an ``undershoot"), the error is $\epsilon_{\rho,n} = 0.5 - 0.3 = 0.2$.
\end{itemize}
This example demonstrates that deviations in the wrong direction are penalized more heavily. Although both 0.3 and 0.7 are 0.2 away from 0.5, 0.7 lies closer to the boundaries of the target interval, so its penalty is smaller. \cite{Tang2025} concluded that the Probability of an Imbalance in the Wrong Direction (PIWD), as suggested by \cite{Thall2015}, is not adequate to inform the burn-in length because it fails to capture the magnitude of the impact on the final allocation. Our Expected Final Allocation Error ($\epsilon_{\rho}$) incorporates this idea, penalizing allocations towards the wrong arm based on the actual number of patients, not just the probability of an error.

\paragraph{Impact on burn-in}
A high Expected Final Allocation Error $\epsilon_{\rho}$ suggests that the design is frequently ``confused", exhibiting high bias or instability, and failing to converge within the prudent target interval. This confusion can even lead to the algorithm allocating a substantial proportion of patients to the wrong arm.
This indicates a clear need for a larger burn-in. This ``protects" the trial from being misled by early, random data, thereby reducing the risk of high bias and lowering the final expected error $\epsilon_{\rho}$.

\section{Formula for burn-in length}\label{s:method}
To determine an appropriate length for the burn-in period for a specific RAR design in a particular trial with $n$ available patients, we synthesize the metrics introduced in Section \ref{s:metric}. We propose a rule that balances the problem difficulty ($\delta$ and $n$) against the design's specific risks ($r$ and $\epsilon_\rho$) and scales with the trial size ($n$).
We propose using the following rule to determine the number of burn-in patients $b$ per arm:
\begin{equation}
    b = \max\left\{ 2,\;
    \left\lfloor
    0.5 \cdot \frac{n \cdot n_{1/2}}{n+n_{1/2}} \cdot (r+\epsilon_\rho)^{\delta}
    \right\rfloor
    \right\}
    \label{eq:burnin_formula}
\end{equation}
where the formula is floored at 2, representing the smallest burn-in. The $\lfloor \cdot  \rfloor$ operation is used to round down the calculated value to the nearest integer. The components are:
\begin{itemize}
    \item $0.5 \cdot \frac{n \cdot n_{1/2}}{n+n_{1/2}} \in [1,\text{min} \{n,n_{1/2} \} )$ is the available burn-in budget, where 
    \begin{itemize}
        \item we multiply with 0.5 because we split the budget among 2 arms, 
        \item \(n \in [1,\infty) \) is the total estimated trial size,
        \item and \(n_{1/2} \in [1,\infty)\) is the saturation parameter from Section \ref{sec:samplesize}, which makes the influence of the trial size non-linear.
    \end{itemize}
    
    \item $r \in [0,1/2]$ is the reactiveness parameter (design speed) from Section \ref{sec.reactiveness}.
    \item $\epsilon_\rho \in [0,1/2]$ is the expected final allocation error (design bias/error) from Section \ref{sec:finalerror}.
    \item $\delta \in [0,\infty)$ is the standardized treatment effect (problem difficulty) from Section \ref{s.standardisedTE}.
\end{itemize}

\paragraph{Interaction of components}
The Formula \eqref{eq:burnin_formula} captures the interplay between these factors. The total burn-in $2b$ is a product of three conceptual parts:
\begin{enumerate}
    \item \textbf{Available Sample Size:} The term $\frac{n \cdot n_{1/2}}{n+n_{1/2}}$ serves as the ``burn-in budget". As established in Section \ref{sec:samplesize}, this term scales non-linearly with $n$, ensuring that the burn-in proportion ($2b/n$) decreases for very large trials.

    \item \textbf{Design Risk:} The term $(r+\epsilon_\rho) \in [0,1]$ is the ``risk multiplier". It combines the risk of a design that converges to fast to an extreme limit ($r$) with the risk of a design that is biased ($\epsilon_\rho$). A perfectly stable and ``non-adaptive" design (like ER) would have $r=0$ and $\epsilon_\rho=0$, causing the formula to default to the smallest burn-in of 2.

    \item \textbf{Problem Difficulty:} The standardized treatment effect $\delta$ acts as an \textit{exponent} on the Design Risk term. This creates the most important interaction:
    \begin{itemize}
        \item When the problem is difficult (i.e., $\delta$ is small, close to 0), the exponent is small. Since $(r+\epsilon_\rho)$ is a fraction in $[0,1]$, a small exponent inflates this risk term (e.g., $0.5^{0.3} \approx 0.81$). This significantly increases the burn-in, reflecting the high need for caution.
        \item When the problem is easy (i.e., $\delta$ is large), the exponent is large. This shrinks the risk term (e.g., $0.5^{2.0} = 0.25$). This reduces the burn-in, as the strong signal (high $\delta$) means even a risky design (high $r$) is less dangerous and will likely converge correctly.
    \end{itemize}
\end{enumerate}
In summary, the rule first establishes a burn-in budget based on trial size, then multiplies it by a risk factor derived from the design's reactiveness and instability. This result is then critically amplified or dampened based on the difficulty of the specific clinical problem, as captured by $\delta$.

For any given RAR design and trial, one can estimate the reactiveness parameter $r$, the expected final allocation error $\epsilon_\rho$, and the standardized treatment effect $\delta$ to determine the appropriate burn-in. This framework is general, as these metrics can be estimated for any RA(R) procedure, including those not investigated in this publication or novel designs yet to be developed.

\section{Simulation studies}\label{s:simu}

We present two examples with differing problem difficulty ($\delta$) and sample size ($n$) to illustrate that adaptive designs are not “off-the-shelf” methods—they require careful adjustment. Our burn-in formula adapts to each scenario, yielding distinct results for the early-phase and late-phase examples.

\subsection{Metrics affected by the burn-in}\label{s.metricaffectedBurnin}

The burn-in length affects operating characteristics such as
\begin{itemize}
\item \textbf{Type-I error rate:} The probability of incorrectly rejecting the null hypothesis $H_0: p_0 = p_1$ when it is true. 

\item \textbf{Statistical power:} The probability of correctly rejecting $H_0$ when a specific alternative hypothesis $H_1: p_0 \neq p_1$ is true.

\item \textbf{Proportion of patients allocated to the best arm:} A measure of in-trial patient benefit, calculated as 
\begin{equation*}
    \text{max \{}E[n_{0}(n) / n \mid p_0 \geq p_1], E[n_{1}(n) / n \mid p_1 >p_0] \}.
\end{equation*}

\item \textbf{Mean Squared Error (MSE):} A measure of estimation accuracy for the treatment effect $\Delta = p_1 - p_0$. It is the expected squared difference between the estimator ($\hat{\Delta} = \hat{p}_1 - \hat{p}_0$) and its true value: $E[(\hat{\Delta} - \Delta)^2 \mid p_0, p_1]$.
\end{itemize} 

RAR designs, particularly those targeting optimal proportions or BRAR, can suffer from substantial type-I error rate inflation when no burn-in period is used \citep{pin2025revisitingoptimalallocationsbinary, Tang2025}.
\cite{Tang2025} specifically investigates the role of the burn-in length, showing that increasing the burn-in reduces this type-I error inflation, but does not fully mitigate it. Furthermore, they demonstrate that the burn-in length has a substantial influence on both statistical power and participant benefit (such as the \% of patients allocated to the best arm). Crucially, these operating characteristics are often not maximized at the minimum or maximum possible burn-in length, suggesting a complex trade-off where an intermediate burn-in duration may be optimal.

We will investigate our proposed formula against three fixed burn-in baselines: the minimal burn-in, $b=2$; a ``mid-point" fixed burn-in, $b = n/3$; and the maximal burn-in, $b=n/2$, which corresponds to a full ER design. The $b=2$ and $b=n/2$ choices represent the most extreme (and common) arbitrary selections. We include the $b \approx 0.33n$ comparator as s a fixed alternative, inspired by literature on optimal timing for other adaptive trials (e.g., early stopping and sample size re-estimation often place the interim analysis between one-third and one-half of the total sample size) \citep{Togo2013, Ohrn2011}. Our simulation aims to demonstrate that our formula achieves a better balance of operating characteristics than any of these fixed choices. 

\subsection{Statistical tests for final analysis}

For the final analysis of the trial, we test the null hypothesis $H_0: p_0 = p_1$ against the two-sided alternative $H_1: p_0 \neq p_1$ at a significance level $\alpha = 0.05$. Let $n_k$ be the total number of patients allocated to arm $k \in \{0, 1\}$, and let $S_k$ be the number of successes observed on arm $k$. The success probability on each arm is estimated as $\hat{p}_k = S_k / n_k$.

\subsubsection{Wald test}
The (unpooled) Wald test is a standard test for this hypothesis. The test statistic is given by:
\begin{equation}
    Z_1 = \frac{\hat{p}_1 - \hat{p}_0}{\sqrt{\frac{\hat{p}_0(1-\hat{p}_0)}{n_0} + \frac{\hat{p}_1(1-\hat{p}_1)}{n_1}}}
    \label{eq:wald_test}
\end{equation}
$H_0$ is rejected if $|Z_1| > z_{1-\alpha/2}$, where $z_{1-\alpha/2}$ is the $1-\alpha/2$ quantile of the standard normal distribution.

\subsubsection{Score test}
The (pooled) score test is often recommended for its more robust performance, particularly in RAR designs, as it maintains the nominal type-I error rate more effectively \citep{Eberhardt1977}. It is derived under the null hypothesis $H_0$, assuming a common success probability $p$. This common probability is estimated by the pooled estimator:
\begin{equation}
    \bar{p} = \frac{S_0 + S_1}{n_0 + n_1} = \frac{S_0 + S_1}{n}
\end{equation}
The score test statistic is then:
\begin{equation}
    Z_0 = \frac{\hat{p}_1 - \hat{p}_0}{\sqrt{\bar{p}(1-\bar{p})\left(\frac{1}{n_0} + \frac{1}{n_1}\right)}}
    \label{eq:score_test}
\end{equation}
$H_0$ is rejected if $|Z_0| > z_{1-\alpha/2}$.

\subsection{ARREST trial}

Our first case study is based on the parameters from the ARREST (Advanced R$^2$Eperfusion STrategies for Refractory Cardiac Arrest) trial \citep{yannopoulos2020advanced}. This trial investigated a novel ECMO-facilitated resuscitation (experimental arm) against standard advanced cardiac life support (control arm) for adults experiencing refractory out-of-hospital cardiac arrest. The primary endpoint was survival to hospital discharge.

We adopt the parameters from the trial's alternative hypothesis for our simulations: a control success rate of $p_0=0.12$ and an experimental success rate of $p_1=0.37$. The total sample size is set to $n=86$ to yield approximately 80\% power under ER for the Wald test. This scenario represents a case with a large absolute treatment difference ($\Delta = 0.25$) and a moderate standardized treatment effect ($\delta \approx 0.3095$).

\subsubsection{Burn-in considerations}
The sample size budget for the burn-in is $\approx 85.27$, standardized treatment effect is $\delta\approx0.3095$ and the values for the reactiveness $r$ and expected final allocation error $\epsilon_\rho$ are displayed in Table \ref{tab:AE_ARREST} for the different designs alongside the estimated recommended burn-in $b$.

The PBB design acts as a benchmark with a burn-in of $b=32$, corresponding to a $BP$ of $72.66\%$. For the adaptive designs evaluated, the recommended burn-in $b$ varies significantly, ranging from a minimum of 12 for BRAR (T) to a maximum of 27 for $R_1$. PTW is the most reactive ($r = 31.95$), apart from the PBB design, but notably maintains a low allocation error ($\epsilon_{\rho} = 0.96$). In contrast, $R_1$ is only moderately reactive ($r = 13.66$) but, together with $N_1$, it incurs the highest expected final allocation error ($\epsilon_{\rho} = 29.06$ and $\epsilon_{\rho} = 32.86$, respectively). This high error results in $R_1$ having the largest combined $r + \epsilon_{\rho}$ score (42.73) among all tested designs and therefore the largest burn-in.

\begin{table}[]
    \centering
    \caption{Estimated reactiveness parameters and burn-in recommendations for the ARREST trial where~$p_0=0.12$, $p_1=0.37$ and~$n=86$ and CALISTO trial  where~$p_0=0.941$, $p_1=0.991$ and~$n=360$ (1000 simulations). CIR stands for confidence interval radius~(95\%, based on a normal approximation).}
    \small
\begin{tabular}{lrrrrr}
\hline
\multicolumn{6}{c}{ARREST} \\
\hline
Design  & r (x100) & $\epsilon_{\rho}$ (x100) & $r + \epsilon_{\rho}$ (x100) & $b$ & \multicolumn{1}{c}{$BP$ (x100)} \\ 
\hline
\nopagebreak ER  & 0.00 & 0.00 & 0.00 & - & - \\
\nopagebreak PBB  & 57.60 & 0.00 & 57.60 & 32 & 72.66 \\
\nopagebreak BRAR (U)  & 22.19 & 0.16 & 22.36 & 20 & 46.40 \\
\nopagebreak BRAR (T)  & 6.62 & 0.06 & 6.68 & 12 & 27.12 \\
\nopagebreak $N_0$  & 27.94 & 2.05 & 29.99 & 23 & 52.63 \\
\nopagebreak $N_1$  & 4.83 & 32.86 & 37.69 & 26 & 59.02 \\
\nopagebreak $R_0$  & 19.65 & 1.76 & 21.40 & 20 & 45.61 \\
\nopagebreak $R_1$  & 13.66 & 29.06 & 42.73 & 27 & 62.27 \\
\nopagebreak PTW  & 31.95 & 0.96 & 32.91 & 24 & 54.64 \\
\nopagebreak RPW  & 18.82 & 1.44 & 20.26 & 18 & 41.70 \\\hline
\nopagebreak max CIR  & 1.01 & 0.89 & 0.96 & 0.49 & 1.13 \\
\hline \\
\hline
\multicolumn{6}{c}{CALISTO} \\
\hline
Design  & r (x100) & $\epsilon_{\rho}$ (x100) & $r + \epsilon_{\rho}$ (x100) & $b$ & \multicolumn{1}{c}{$BP$ (x100)} \\ 
\hline
\nopagebreak ER  & 0.00 & 0.00 & 0.00 & - & - \\
\nopagebreak PBB  & 70.03 & 0.00 & 70.03 & 124 & 68.55 \\
\nopagebreak BRAR (U)  & 25.64 & 0.51 & 26.15 & 98 & 54.09 \\
\nopagebreak BRAR (T)  & 4.13 & 0.02 & 4.15 & 69 & 37.87 \\
\nopagebreak $N_0$  & 22.00 & 4.11 & 26.12 & 98 & 54.38 \\
\nopagebreak $N_1$  & 24.83 & 27.04 & 51.87 & 116 & 63.95 \\
\nopagebreak $R_0$  & 7.26 & 0.21 & 7.47 & 61 & 33.44 \\
\nopagebreak $R_1$  & 7.42 & 0.21 & 7.64 & 67 & 36.84 \\
\nopagebreak PTW  & 35.10 & 2.77 & 37.86 & 106 & 58.80 \\
\nopagebreak RPW  & 7.56 & 5.66 & 13.22 & 79 & 43.76 \\\hline
\nopagebreak max CIR  & 1.05 & 0.53 & 1.15 & 2.03 & 1.13 \\
\hline 
\end{tabular}

    \label{tab:AE_ARREST}
\end{table}

\subsubsection{Simulations results}
The simulation results for the ARREST trial, with ER serving as the baseline (type-I error rate $= 5.93\%$ for $Z_1$ and $Z_0$, Power $\approx 80\%$, MSE $= 0.0078$), reveal the critical importance of the burn-in period.

The most striking observation is the failure of highly reactive designs when implemented with the smallest burn-in of $b=2$. Designs such as PBB, $N_1$, $R_1$, and BRAR (U) exhibit extreme type-I error rate inflation for $Z_1$ (e.g., $78.24\%$, $89.80\%$, $89.65\%$, and $20.86\%$, respectively) - as pointed out by \cite{pin2025revisitingoptimalallocationsbinary} for $N_1$ and $R_1$- and high MSE. This demonstrates that immediate adaptation is nonviable for these methods.

Implementing a larger burn-in is essential. The comparison between our flexible $b$ (e.g., $b=26$ for $N_1$, $b=32$ for PBB) and the fixed $b \approx n/3$ baseline ($b=29$) is nuanced. The $b=29$ rule is a strong performer for controlling type-I error rate ($Z_1$) and power ($Z_0$) in many designs. However, our flexible $b$ formula often finds a better balance for the Score test, securing better type-I error rate ($Z_0$) control in several key designs (PBB, BRAR(U), BRAR(T), RPW).

For the BRAR designs, our flexible $b$ ($b=20$ and $b=12$) offers the best type-I error rate ($Z_0$) control (2.70\% and 4.21\%, respectively). The PTW design emerges as a special case where $b=2$ is surprisingly effective, securing the best type-I error rate for both $Z_1$ (5.12\%) and $Z_0$ (4.51\%), as well as the best MSE (0.0075). For RPW, our flexible $b$ ($b=18$) provides a good balance, with optimal type-I error rate ($Z_0$) (4.52\%) and power ($Z_1$) (81.18\%).

In general, increasing the burn-in period from $b=2$ to either the flexible $b$ or the $b \approx n/3$ fixed value consistently maintains or improves MSE and leads to better type-I error rate control.

\begin{table}[]
    \centering
    \caption{ARREST trial where~$p_0=0.12$, $p_1=0.37$ and~$n=86$ (10000 simulations). \
    \textbf{Color Scheme:} Compares each value to the baseline \textbf{ER} row or a fixed target. \
    \textbf{Type-I Error} ($Z_1, Z_0$) compares to an ideal 5 (Dark Green: 4--5, Light Green: 0--4 or 5--6, Yellow: 6--10, Red: $>$10). \
    \textbf{Power} ($Z_1, Z_0$) compares to ER values (80.88, 79.94) (Dark Green: $>$ER, Light Green: within 2 points, Yellow: 2--5 points below, Red: $>$5 points below). \
    \textbf{$n_1/n$} compares to 0.500 (Yellow: 0.475--0.525, Dark Green: $>$0.525, Red: $<$0.475). \
    \textbf{MSE} compares to ER (0.0078) $\pm$0.001 (Dark Green: 0.0070--0.0090, Red: $>$0.0090, Yellow: $<$0.0070).
    \newline 
    \textbf{Bolding:} For a given design, indicates the \textbf{best} value among the three burn-in options (b=2, flexible b, b=N/3). ``Best" is defined as: type-I error rate  closest to 5.0 within [0, 5.2], Power highest, $n_1/n$ highest, and MSE lowest.
    }
\begin{tabular}{llrrrrrr}
\hline
Design & Burn-In & \multicolumn{2}{c}{Type-I error} & \multicolumn{2}{c}{Power} & $n_1/n$ & MSE \\
& & $Z_1$ & $Z_0$ & $Z_1$ & $Z_0$ & & \\
\hline
ER         & -       & \cellcolor{myLightGreen}5.93 & \cellcolor{myLightGreen}5.93 & \cellcolor{myLightGreen}80.88 & \cellcolor{myLightGreen}79.94 & \cellcolor{myYellow}0.500 & \cellcolor{myDarkGreen}0.0078 \\
PBB        & 2       & \cellcolor{myRed}78.24 & \cellcolor{myLightGreen}5.83 & \cellcolor{myYellow}78.47 & \cellcolor{myRed}0.42 & \cellcolor{myDarkGreen}\textbf{0.977} & \cellcolor{myRed}0.0591 \\
           & 32      & \cellcolor{myYellow}6.66  & \cellcolor{myDarkGreen}\textbf{4.17} & \cellcolor{myLightGreen}\textbf{79.56} & \cellcolor{myRed}74.34 & \cellcolor{myDarkGreen}0.628 & \cellcolor{myDarkGreen}\textbf{0.0078} \\
           & 29      & \cellcolor{myDarkGreen}\textbf{4.82} & \cellcolor{myYellow}7.06 & \cellcolor{myRed}71.86 & \cellcolor{myLightGreen}\textbf{79.71} & \cellcolor{myDarkGreen}0.663 & \cellcolor{myDarkGreen}0.0078 \\
BRAR (U)   & 2       & \cellcolor{myRed}20.86 & \cellcolor{myLightGreen}0.64 & \cellcolor{myRed}74.82 & \cellcolor{myRed}40.71 & \cellcolor{myDarkGreen}\textbf{0.835} & \cellcolor{myRed}0.0113 \\
           & 20      & \cellcolor{myRed}13.83 & \cellcolor{myLightGreen}\textbf{2.70}  & \cellcolor{myYellow}\textbf{77.97} & \cellcolor{myRed}66.21 & \cellcolor{myDarkGreen}0.735 & \cellcolor{myDarkGreen}0.0079 \\
           & 29      & \cellcolor{myDarkGreen}\textbf{4.38} & \cellcolor{myYellow}9.90 & \cellcolor{myRed}73.80 & \cellcolor{myDarkGreen}\textbf{80.14} & \cellcolor{myDarkGreen}0.649 & \cellcolor{myDarkGreen}\textbf{0.0078} \\
BRAR (T)   & 2       & \cellcolor{myYellow}9.95 & \cellcolor{myDarkGreen}4.10 & \cellcolor{myLightGreen}\textbf{80.78} & \cellcolor{myRed}74.03 & \cellcolor{myDarkGreen}\textbf{0.691} & \cellcolor{myDarkGreen}0.0081 \\
           & 12      & \cellcolor{myYellow}9.19  & \cellcolor{myDarkGreen}\textbf{4.21}  & \cellcolor{myLightGreen}80.31 & \cellcolor{myRed}73.79 & \cellcolor{myDarkGreen}0.685 & \cellcolor{myDarkGreen}0.0080 \\
           & 29      & \cellcolor{myDarkGreen}\textbf{4.40} & \cellcolor{myYellow}7.67 & \cellcolor{myYellow}76.38 & \cellcolor{myDarkGreen}\textbf{80.65} & \cellcolor{myDarkGreen}0.615 & \cellcolor{myDarkGreen}\textbf{0.0079} \\
$N_0$      & 2       & - & \cellcolor{myLightGreen}5.94 & - & \cellcolor{myLightGreen}\textbf{79.52} & \cellcolor{myRed}0.393 & \cellcolor{myDarkGreen}0.0090 \\
           & 23      & - & \cellcolor{myLightGreen}5.52  & - & \cellcolor{myLightGreen}79.15 & \cellcolor{myRed}0.399 & \cellcolor{myRed}0.0091 \\
           & 29      & - & \cellcolor{myLightGreen}\textbf{3.59} & - & \cellcolor{myLightGreen}78.27 & \cellcolor{myRed}\textbf{0.415} & \cellcolor{myDarkGreen}\textbf{0.0087} \\
$N_1$      & 2       & \cellcolor{myRed}89.80 & - & \cellcolor{myDarkGreen}\textbf{94.79} & - & \cellcolor{myDarkGreen}\textbf{0.714} & \cellcolor{myRed}0.0530 \\
           & 26      & \cellcolor{myRed}12.16 & - & \cellcolor{myDarkGreen}81.86 & - & \cellcolor{myDarkGreen}0.591 & \cellcolor{myDarkGreen}0.0077 \\
           & 29      & \cellcolor{myDarkGreen}\textbf{4.96} & - & \cellcolor{myLightGreen}79.04 & - & \cellcolor{myDarkGreen}0.581 & \cellcolor{myDarkGreen}\textbf{0.0077} \\
$R_0$      & 2       & - & \cellcolor{myLightGreen}5.50 & - & \cellcolor{myYellow}77.74 & \cellcolor{myRed}0.442 & \cellcolor{myDarkGreen}0.0082 \\
           & 20      & - & \cellcolor{myLightGreen}5.75 & - & \cellcolor{myLightGreen}79.07 & \cellcolor{myRed}0.442 & \cellcolor{myDarkGreen}0.0080 \\
           & 29      & - & \cellcolor{myLightGreen}\textbf{3.93} & - & \cellcolor{myLightGreen}\textbf{79.25} & \cellcolor{myRed}\textbf{0.446} & \cellcolor{myDarkGreen}\textbf{0.0078} \\
$R_1$      & 2       & \cellcolor{myRed}89.65 & - & \cellcolor{myDarkGreen}\textbf{94.84} & - & \cellcolor{myDarkGreen}\textbf{0.753} & \cellcolor{myRed}0.0347 \\
           & 27      & \cellcolor{myRed}11.37 & - & \cellcolor{myDarkGreen}82.07 & - & \cellcolor{myDarkGreen}0.609 & \cellcolor{myDarkGreen}\textbf{0.0075} \\
           & 29      & \cellcolor{myLightGreen}\textbf{5.36} & - & \cellcolor{myYellow}78.42 & - & \cellcolor{myDarkGreen}0.598 & \cellcolor{myDarkGreen}0.0077 \\
PTW        & 2       & \cellcolor{myLightGreen}\textbf{5.12} & \cellcolor{myDarkGreen}\textbf{4.51} & \cellcolor{myDarkGreen}81.01 & \cellcolor{myLightGreen}78.84 & \cellcolor{myDarkGreen}\textbf{0.578} & \cellcolor{myDarkGreen}\textbf{0.0075} \\
           & 24      & \cellcolor{myLightGreen}5.61 & \cellcolor{myDarkGreen}4.41 & \cellcolor{myDarkGreen}\textbf{81.38} & \cellcolor{myLightGreen}79.12 & \cellcolor{myDarkGreen}0.536 & \cellcolor{myDarkGreen}0.0076 \\
           & 29      & \cellcolor{myDarkGreen}4.60 & \cellcolor{myLightGreen}5.87 & \cellcolor{myLightGreen}79.23 & \cellcolor{myDarkGreen}\textbf{81.65} & \cellcolor{myDarkGreen}0.526 & \cellcolor{myDarkGreen}0.0075 \\
RPW        & 2       & \cellcolor{myYellow}6.05 & \cellcolor{myDarkGreen}4.42 & \cellcolor{myLightGreen}80.26 & \cellcolor{myYellow}76.87 & \cellcolor{myDarkGreen}\textbf{0.578} & \cellcolor{myDarkGreen}0.0077 \\
           & 18      & \cellcolor{myLightGreen}5.84  & \cellcolor{myDarkGreen}\textbf{4.52}  & \cellcolor{myDarkGreen}\textbf{81.18} & \cellcolor{myLightGreen}78.45 & \cellcolor{myDarkGreen}0.559 & \cellcolor{myDarkGreen}0.0077 \\
           & 29      & \cellcolor{myDarkGreen}\textbf{4.80} & \cellcolor{myLightGreen}5.66 & \cellcolor{myLightGreen}79.08 & \cellcolor{myDarkGreen}\textbf{81.42} & \cellcolor{myDarkGreen}0.536 & \cellcolor{myDarkGreen}\textbf{0.0077} \\
\hline
\end{tabular}
    \label{tab:ARREST}
\end{table}

\subsection{CALISTO trial}
As a second case study to illustrate the application of the proposed methods, we use the CALISTO trial \citep{Decousus2010}. This study investigated the efficacy of Arixtra (treatment) compared to a placebo in patients with acute symptomatic lower limb thrombophlebitis. The primary efficacy endpoint was a composite measure; a successful outcome was defined as the absence of death, symptomatic pulmonary embolism, symptomatic deep-vein thrombosis, or symptomatic recurrence/extension of the thrombosis by day 47.
The study reported high success rates of 99.1\% for the Arixtra group and 94.1\% for the placebo group. This corresponds to a small absolute treatment difference ($\Delta=0.05$) and, due to the high variance near the boundary, a very small standardized treatment effect ($\delta\approx0.1515$). 

While the original study enrolled 1502 patients (of a planned 3002), we adopt its parameters to redesign a new hypothetical trial. For our simulations, we set the total sample size to $n=360$, as this is the size required to achieve approximately 80\% power with ER.

\subsubsection{Burn-in considerations}
Given a sample size budget of approximatly $347.5$ and a standardized treatment effect of $\delta \approx 0.1515$, the trade-offs between design reactiveness $r$, expected final allocation error $\epsilon_{\rho}$, and the recommended burn-in $b$ are displayed in Table \ref{tab:AE_ARREST}. The PBB design serves as a benchmark with $b=124$ and a $BP$ of $68.55\%$. The recommended burn-in $b$ for the adaptive designs spans from 61 for $R_0$ to 116 for $N_1$. The PTW design is the most reactive ($r = 35.10$) while maintaining a relatively low allocation error ($\epsilon_{\rho} = 2.77$). Conversely, the $N_1$ design, despite having moderate reactiveness ($r = 24.83$), incurs by far the highest expected final allocation error ($\epsilon_{\rho} = 27.04$), resulting in the largest combined $r + \epsilon_{\rho}$ score (51.87) among all designs.

Compared to the ARREST trial, several differences are notable. Due to the larger sample size, the total range of the recommended burn-in $b$ is larger (61--116 vs. 12--27). The largest $BP$, represented by the PBB, has decreased to $68.55\%$ (from $72.66\%$), which can be attributed to non-linear sample size influences. Furthermore, the design requiring the largest burn-in has changed from $R_1$ in ARREST to $N_1$ in this trial. This shift is partially due to the different treatment effect and sample size, but also reflects a change in the relative orders of reactiveness $r$ and error $\epsilon_{\rho}$ across the designs.

\subsubsection{Simulations results}

The CALISTO trial simulations, with an ER baseline showing near-ideal type-I error rate control ($Z_1=4.91\%$, $Z_0 = 4.90\%$) and 79.5\% power, demonstrate that the $b=2$ burn-in is completely nonviable in this setting. It leads to catastrophic $Z_1$ Type-I error inflation (e.g., $96.23\%$ for $N_1$) and massively inflated MSE. 

This establishes that a substantial burn-in is mandatory. The core trade-off for any large burn-in is sacrificing patient benefit ($n_1/n$) to gain type-I error rate control and MSE stability. The $b \approx n/3$ ($b=120$) rule and our flexible $b$ rule both successfully fix the type-I error rate and MSE issues that $b=2$ failed. The crucial difference lies in the efficiency of this trade-off.

The fixed $b=120$ rule is a strong performer for the $Z_1$ (Wald) test, as it is essential for ``rescuing" the $Z_1$ power from its $b=2$ collapse. However, our flexible $b$ formula is overwhelmingly superior for the $Z_0$ (Score) test, securing the best $Z_0$ power in 8 of the 9 adaptive designs. This $b=120$ rule's ``win" on $Z_1$ power comes at a high cost, as it consistently reduces patient benefit ($n_1/n$) and $Z_0$ power compared to the flexible $b$.

For the BRAR designs, our flexible $b$ ($b=98$ and $b=69$) offers the best $Z_0$ power (80.63\% and 81.89\%) and higher patient benefit. The fixed $b=120$ rule, while boosting $Z_1$ power, cuts $Z_0$ power significantly (e.g., to $73.15\%$ for BRAR(T)) and reduces $n_1/n$. For RPW, our flexible $b$ ($b=79$) provides the best overall balance: it has the best $Z_0$ power ($80.12\%$), excellent type-I error rate control for both tests ($Z_1=5.07\%, Z_0=5.24\%$), the best MSE ($0.0004$), and higher patient benefit. The $b=120$ rule's ``win" on $Z_1$ power is trivial ($79.97\%$ vs $79.39\%$) and not worth the cost to other metrics.

In conclusion, increasing the burn-in period from $b=2$ is mandatory. The fixed $b \approx n/3$ rule is a blunt instrument that prioritizes $Z_1$ (Wald) power at the expense of $Z_0$ (Score) power and patient benefit. Our flexible $b$ provides a far superior and more balanced trade-off, especially for the trialist who values patient benefit and intends to use the more robust $Z_0$ test for their final analysis.

\begin{table}[!htbp]
\centering
\caption{CALISTO Trial with $p_0=0.941$, $p_1=0.991$ and~$n=360$ (10000 simulations)
\newline 
\textbf{Color Scheme:} Compares each value to the baseline \textbf{ER} row or a fixed target. \
    \textbf{Type-I Error} ($Z_1, Z_0$) compares to an ideal 5 (Dark Green: 4--5, Light Green: 0--4 or 5--6, Yellow: 6--10, Red: $>$10). \
    \textbf{Power} ($Z_1, Z_0$) compares to the ER values (79.53, 79.53) (Dark Green: $>$ER, Light Green: within 2 points, Yellow: 2--5 points below, Red: $>$5 points below). \
    \textbf{$n_1/n$} compares to 0.500 (Yellow: 0.475--0.525, Dark Green: $>$0.525, Red: $<$0.475). \
    \textbf{MSE} compares to ER (0.0004) $\pm$0.0002 (Dark Green: 0.0002--0.0006, Red: $>$0.0006, Yellow: $<$0.0002).
    \newline 
    \textbf{Bolding:} For a given design, indicates the \textbf{best} value among the three burn-in options (b=2, flexible b, b=N/3). ``Best" is defined as: type-I error rate closest to 5.0 within [0, 5.2], Power highest, $n_1/n$ highest, and MSE lowest.
}
\begin{tabular}{llllllll}
\hline
Design & Burn-In & \multicolumn{2}{c}{Type-I error} & \multicolumn{2}{c}{Power} & $n_1/n$ & MSE \\
& & $Z_1$ & $Z_0$ & $Z_1$ & $Z_0$ & & \\
\hline
ER         & -       & \cellcolor{myDarkGreen}4.91 & \cellcolor{myDarkGreen}4.90 & \cellcolor{myLightGreen}79.53 & \cellcolor{myLightGreen}79.53 & \cellcolor{myYellow}0.500 & \cellcolor{myDarkGreen}0.0004 \\
PBB        & 2       & \cellcolor{myRed}88.58 & \cellcolor{myRed}11.74 & \cellcolor{myRed}35.58 & \cellcolor{myRed}12.15 & \cellcolor{myDarkGreen}\textbf{0.994} & \cellcolor{myRed}0.0289 \\
           & 124     & \cellcolor{myLightGreen}5.45  & \cellcolor{myDarkGreen}\textbf{4.71}  & \cellcolor{myRed}71.10 & \cellcolor{myDarkGreen}\textbf{81.57} & \cellcolor{myDarkGreen}0.656 & \cellcolor{myDarkGreen}0.0005 \\
           & 120     & \cellcolor{myDarkGreen}\textbf{4.79} & \cellcolor{myLightGreen}5.24 & \cellcolor{myDarkGreen}\textbf{79.57} & \cellcolor{myRed}67.91 & \cellcolor{myDarkGreen}0.667 & \cellcolor{myDarkGreen}\textbf{0.0005} \\
BRAR (U)   & 2       & \cellcolor{myLightGreen}2.62 & \cellcolor{myRed}16.94 & \cellcolor{myRed}5.13 & \cellcolor{myYellow}76.12 & \cellcolor{myDarkGreen}\textbf{0.881} & \cellcolor{myRed}0.0176 \\
           & 98      & \cellcolor{myLightGreen}\textbf{2.93}  & \cellcolor{myYellow}6.55  & \cellcolor{myRed}56.24 & \cellcolor{myDarkGreen}\textbf{80.63} & \cellcolor{myDarkGreen}0.708 & \cellcolor{myDarkGreen}0.0005 \\
           & 120     & \cellcolor{myLightGreen}5.94 & \cellcolor{myLightGreen}\textbf{3.07} & \cellcolor{myDarkGreen}\textbf{82.80} & \cellcolor{myRed}71.10 & \cellcolor{myDarkGreen}0.655 & \cellcolor{myDarkGreen}\textbf{0.0005} \\
BRAR (T)   & 2       & \cellcolor{myLightGreen}\textbf{3.38} & \cellcolor{myYellow}6.39 & \cellcolor{myRed}63.66 & \cellcolor{myDarkGreen}81.21 & \cellcolor{myDarkGreen}\textbf{0.701} & \cellcolor{myRed}0.0007 \\
           & 69      & \cellcolor{myLightGreen}3.31 & \cellcolor{myYellow}6.02  & \cellcolor{myRed}67.31 & \cellcolor{myDarkGreen}\textbf{81.89} & \cellcolor{myDarkGreen}0.681 & \cellcolor{myDarkGreen}0.0006 \\
           & 120     & \cellcolor{myLightGreen}5.29 & \cellcolor{myLightGreen}\textbf{3.70} & \cellcolor{myDarkGreen}\textbf{82.12} & \cellcolor{myRed}73.15 & \cellcolor{myDarkGreen}0.620 & \cellcolor{myDarkGreen}\textbf{0.0005} \\
$N_0$      & 2       & - & \cellcolor{myLightGreen}5.25 & - & \cellcolor{myDarkGreen}\textbf{79.97} & \cellcolor{myDarkGreen}\textbf{0.723} & \cellcolor{myRed}0.0016 \\
           & 98      & - & \cellcolor{myLightGreen}\textbf{5.17}  & - & \cellcolor{myDarkGreen}79.86 & \cellcolor{myDarkGreen}0.663 & \cellcolor{myDarkGreen}0.0005 \\
           & 120     & - & \cellcolor{myLightGreen}3.51 & - & \cellcolor{myRed}70.65 & \cellcolor{myDarkGreen}0.624 & \cellcolor{myDarkGreen}\textbf{0.0004} \\
$N_1$      & 2       & \cellcolor{myRed}96.23 & - & \cellcolor{myDarkGreen}\textbf{94.17} & - & \cellcolor{myRed}0.158 & \cellcolor{myRed}0.0009 \\
           & 116     & \cellcolor{myYellow}7.92  & - & \cellcolor{myDarkGreen}83.94 & - & \cellcolor{myRed}0.363 & \cellcolor{myDarkGreen}\textbf{0.0003} \\
           & 120     & \cellcolor{myDarkGreen}\textbf{4.80} & - & \cellcolor{myYellow}75.50 & - & \cellcolor{myRed}\textbf{0.371} & \cellcolor{myDarkGreen}0.0003 \\
$R_0$      & 2       & - & \cellcolor{myLightGreen}5.75 & - & \cellcolor{myDarkGreen}81.37 & \cellcolor{myDarkGreen}\textbf{0.672} & \cellcolor{myRed}0.0007 \\
           & 61      & - & \cellcolor{myYellow}6.32 & - & \cellcolor{myDarkGreen}\textbf{81.39} & \cellcolor{myDarkGreen}0.661 & \cellcolor{myDarkGreen}0.0005 \\
           & 120     & - & \cellcolor{myLightGreen}\textbf{3.48} & - & \cellcolor{myRed}71.09 & \cellcolor{myDarkGreen}0.599 & \cellcolor{myDarkGreen}\textbf{0.0004} \\
$R_1$      & 2       & \cellcolor{myLightGreen}5.65 & - & \cellcolor{myLightGreen}78.79 & - & \cellcolor{myYellow}\textbf{0.506} & \cellcolor{myRed}0.0024 \\
           & 67      & \cellcolor{myLightGreen}5.28  & - & \cellcolor{myLightGreen}79.05 & - & \cellcolor{myYellow}0.505 & \cellcolor{myDarkGreen}0.0004 \\
           & 120     & \cellcolor{myDarkGreen}\textbf{4.94} & - & \cellcolor{myLightGreen}\textbf{79.45} & - & \cellcolor{myYellow}0.505 & \cellcolor{myDarkGreen}\textbf{0.0004} \\
PTW        & 2       & \cellcolor{myLightGreen}2.72 & \cellcolor{myYellow}6.12 & \cellcolor{myRed}9.56 & \cellcolor{myYellow}74.53 & \cellcolor{myDarkGreen}\textbf{0.847} & \cellcolor{myRed}0.0049 \\
           & 106     & \cellcolor{myDarkGreen}\textbf{4.23}  & \cellcolor{myLightGreen}5.52  & \cellcolor{myRed}69.23 & \cellcolor{myDarkGreen}\textbf{80.31} & \cellcolor{myDarkGreen}0.636 & \cellcolor{myDarkGreen}0.0005 \\
           & 120     & \cellcolor{myLightGreen}5.40 & \cellcolor{myDarkGreen}\textbf{4.41} & \cellcolor{myDarkGreen}\textbf{80.30} & \cellcolor{myRed}72.88 & \cellcolor{myDarkGreen}0.607 & \cellcolor{myDarkGreen}\textbf{0.0004} \\
RPW        & 2       & \cellcolor{myDarkGreen}\textbf{4.85} & \cellcolor{myLightGreen}5.43 & \cellcolor{myRed}67.77 & \cellcolor{myYellow}75.95 & \cellcolor{myDarkGreen}\textbf{0.582} & \cellcolor{myDarkGreen}0.0006 \\
           & 79      & \cellcolor{myLightGreen}5.07  & \cellcolor{myLightGreen}\textbf{5.24}  & \cellcolor{myLightGreen}79.39 & \cellcolor{myDarkGreen}\textbf{80.12} & \cellcolor{myYellow}0.519 & \cellcolor{myDarkGreen}0.0004 \\
           & 120     & \cellcolor{myLightGreen}5.28 & \cellcolor{myLightGreen}5.38 & \cellcolor{myDarkGreen}\textbf{79.97} & \cellcolor{myLightGreen}79.45 & \cellcolor{myYellow}0.510 & \cellcolor{myDarkGreen}\textbf{0.0004} \\
\hline
\end{tabular}
    \label{tab:CALISTO}
\end{table}

\section{Discussion}\label{s:discuss}
This paper introduces the first systematic framework for determining the burn-in length in RAR trials. Moving beyond arbitrary justifications, we explicitly incorporate key factors governing the burn-in decision: the non-linear impact of sample size (Section \ref{sec:samplesize}), two novel metrics to characterize a design's behavior: its reactiveness ($r$) (Section \ref{sec.reactiveness}) and its expected final allocation error ($\epsilon_\rho$) (Section \ref{sec:finalerror}) and the problem's difficulty via the standardized treatment effect ($\delta$). The result of this framework is the principled, general-purpose formula (Equation \eqref{eq:burnin_formula}) that combines these components in a meaningful, interactive way.


Simulation studies (Section \ref{s:simu}) provide a clear and expected warning: implementing highly reactive designs ($N_1$, $R_1$, PBB) with the smallest burn-in ($b=2$) results in severe performance issues, including type-I error rate inflation and inflated MSE. This demonstrates that insufficient initial exploration makes these algorithms vulnerable to misleading early data, leading to statistically unsound results, particularly in challenging scenarios, e.g small true effects (CALISTO) or high variability.

The primary finding is that  our recommended burn-in $b$ acts as an effective stabilizer, it consistently and effectively mitigated the severe type-I error inflation and brought the MSE back in line with the non-adaptive ER baseline. We observed that while the score test $Z_0$ generally provides better type-I error control than the Wald test $Z_1$, a larger burn-in can lead to improved type-I error rate control for both tests across certain designs.

This stabilization must be compared against reasonable heuristics, such as a fixed $b \approx n/3$ rule. Our simulations revealed this fixed approach 
acts as a ``blunt instrument", enforcing a long ER period. While this successfully stabilizes MSE and type-I error rate, and rescued the Wald test's power in challenging settings like CALISTO.
This ``win" came at a high and often unnecessary cost: reduced patient benefit and, critically, degraded power for the more robust score test ($Z_0$).
This is where our flexible formula demonstrates its advantage. It performs an efficient balance, not just stabilization. For example, in the RPW design in CALISTO, our formula found a "sweet spot" missed by fixed rules. It achieved near-optimal type-I error rate, MSE, and $Z_0$ power, whereas the $b \approx n/3$ rule's trivial gains in $Z_1$ power were achieved at the expense of all other key metrics. Our framework thus provides a more nuanced tool that performs robustly across both test statistics while better preserving the patient-benefit and efficiency benefits of adaptation.

The introduction of the novel metrics, reactiveness ($r$) and expected final allocation error ($\epsilon_\rho$), opens new avenues for future research. This framework now allows the first quantitative comparison of the inherent speed and stability of any two-arm RAR algorithm for binary outcomes, applicable even to future or specialized designs. More broadly, the core insights derived here are valuable for other common adaptive designs, like those involving sample size re-estimation or early stopping. Since these decisions are similarly vulnerable to early data variance, understanding the "reactiveness" of their stopping or re-estimation rules is crucial. Applying a principled stability analysis, analogous to the one developed here, could help ensure these adaptations are triggered based on reliable information, but not too late.

A key practical recommendation for future research is the concept of \emph{burn-in re-estimation}. While our simulations used a fixed burn-in $b$, in practice, clinicians could update the estimates of the standardized treatment effect $\delta$ using Maximum Likelihood Estimates (MLEs) from the accumulated data (Equation \eqref{eq:standardTreatDiff}) at the end of the initial burn-in phase. If the estimated effect is smaller or the variability is higher than expected (indicating a more challenging scenario), extending the burn-in period would be warranted before proceeding to the adaptive phase. No adjustment would be needed if the estimates align with prior assumptions. 

This research contains several limitations that open avenues for future work. First, our formula contains a ``meta-parameter", $n_{1/2}$, which we set to 1000 to define a reasonable saturation curve (see Figure \ref{fig:2b_curve}). Future work could explore the formula's sensitivity to this choice and allow practitioners to tune its cautiousness based on risk tolerance. 
Moreover, practitioners may wish to tune the formula's cautiousness based on their risk tolerance. This could be done by scaling the final recommendation (e.g., $b' = C \times b$ for $C=1.2$) or by scaling the exponent (e.g., $b' \propto (r+\epsilon_\rho)^{C \cdot \delta}$) to explicitly modify the formula's sensitivity to the problem's difficulty.
Second, our framework is currently specified for two-arm trials. A non-trivial but necessary extension would be to adapt it for multi-arm ($K>$2) trials. This would likely involve generalizing the $0.5$ scaling factor to $1/K$ and, more complexly, redefining $\delta$ based on a global null hypothesis and adapting $r$ and $\epsilon_\rho$ to handle a $K$-dimensional allocation vector.
Third, for response-adaptive clinical trials with a blocked structure (where the adaptive allocation probability is fixed within blocks \citep{proper2021alternative}), we recommend using Equation \eqref{eq:burnin_formula}.The parameters $r$ and $\epsilon_{\rho}$ should be calculated based on the full RAR procedure, including the specific block randomization. Practically, we suggest setting the first block size equal to the total recommended burn-in length or ensuring the first few blocks maintain a 1:1 randomization ratio until the total participant count covers the recommended burn-in length. Future research could investigate this decision rule's performance in blocked designs.
Fourth, one could explore how other measures of interest, such as the relative risk or odds ratio, affect the burn-in Formula \eqref{eq:burnin_formula}. \cite{Pin_Deming2024} derived how optimal designs change for different measures of interest. Analogously, we would need to redefine the standardized treatment effect $\delta$ based on the measure of interest and transformed arm response variances. 
Fifth, our formula is a principled heuristic designed for practical application. A more complex alternative would be to frame the choice of $b$ as a formal optimization problem. This would involve defining a utility function that weights all key operating characteristics: Power, Patient Benefit ($n_{\text{best}}/n$), type-I Error, and MSE. One could then use extensive simulations to find the $b$ that maximizes this utility. This is a computationally extensive task. Our method is lower-computation in comparison, as it only requires estimating $r$ and $\epsilon_\rho$ to arrive at a direct recommendation, which can serve as a starting point for any finer-grained search. As an alternative to a single utility function, a two-stage optimization may better reflect regulatory priorities: (i) identify the range of $b$ that satisfies hard constraints (e.g., type-I error $\le 0.05$ and MSE $\le \text{MSE}_{\text{ER}}$), and (ii) find the $b$ within that valid range that maximizes a weighted sum of power and patient benefit.
Finally, our work does not consider the impact of delayed outcomes or cohort-based enrollment, which are common practical challenges. Further research could explore how these factors interact with the burn-in phase. 

In conclusion, this paper challenges the long-standing practice of selecting burn-in lengths by ``guesswork". We provide a practical, data-driven, and generalizable tool to move this decision from ad-hoc art to a principled, scientific choice, ultimately fostering the design of safer, more reliable, and more efficient adaptive trials.

\section*{Acknowledgements}
The authors acknowledge the use of large language models (LLMs) to assist with generating figures and refining grammar and wording in this paper. The LLMs were not used for data analysis, interpretation, or original scientific writing. All content has been carefully reviewed and verified by the authors, who take full responsibility for the integrity and accuracy of the work.

The authors acknowledge funding and support from the UK Medical Research Council (grants MC UU 00002/19 (GC), MC UU 00002/15 and MC UU 00040/03 (SSV, DSR, SB)), as well as an MRC Biostatistics Unit Core Studentship (LP) and the Cusanuswerk e.V. (LP). SSV is part of PhaseV's advisory board.

\bibliographystyle{abbrvnat} 
\bibliography{references}

\end{document}